\documentstyle[12pt,psfig]{article}
\advance\textheight by 60pt
\advance\voffset by -40pt
\advance\textwidth by 50pt
\advance\oddsidemargin by -25pt
\advance\evensidemargin by -25pt

\def\single_space{\baselineskip 12pt plus 1pt minus 1pt}
\def\one_and_a_half_space{\baselineskip 19pt plus 1pt minus 1pt}
\def\double_spacesp{\baselineskip 25pt plus 2pt minus 2pt}

\begin{document}
\begin{titlepage}
\begin{flushright}
{\bf October 1999} \\
{\bf PSU/TH-218} 
\end{flushright}
\vskip 1.5cm
\double_spacesp
{\Large
{\bf
\begin{center}
Extra-dimensional gravity and dijet production \\
\vskip 0.1cm
at $\gamma \gamma$ colliders
\end{center}
}}
\vskip 1.0cm
\begin{center}
M.~A.~Doncheski \\
Department of Physics \\
The Pennsylvania State University\\
Mont Alto, PA 17237  USA \\
\vskip 0.1cm
and \\
\vskip 0.1cm
R. W. Robinett \\
Department of Physics\\
The Pennsylvania State University\\
University Park, PA 16802 USA \\
\end{center}
\vskip 1.0cm
\begin{abstract}
In this note, we consider dijet production at $\gamma \gamma$ colliders as a 
probe of recently proposed, large extra-dimensional gravity models.  The 
exchange of virtual, spin-2 graviton towers (Kaluza-Klein excitations) 
significantly modifies the cross section, as compared to the Standard Model 
predictions.  We find that, in order to maximize the value of the effective 
scale that can be probed at a given center-of-mass energy, a very severe 
$p_{_T}\!$ cut should be applied; in general, a $p_{_T}\!$ equal to 
approximately $46\%$ of the $e^+ e^-$ beam energy gives the highest reach.  We 
find that we can probe the effective mass scale from about $2.7 \; TeV$ to 
$11.1 \; TeV$, depending on the center-of-mass energy and assumptions about 
the model.

\end{abstract}
\end{titlepage}
\double_spacesp

\begin{flushleft}
 {\large {\bf I.~Introduction}}
\end{flushleft}
\vskip 0.5cm

A recently proposed model suggests~\cite{ADD1} that gravitational interactions 
take place in $4+n$ dimensions, where the extra $n$-dimensions are large 
({\it i.e.,} as large as millimeter scale) spatial dimensions, commonly 
referred to as the bulk.  Interactions other than gravity (electroweak and 
strong) are confined to the 3-dimensional brane, commonly referred to as the 
wall, which corresponds to the usual 3 spatial dimensions.  The gravitational 
interaction is then understood as appearing to be weak, as we only observe its 
projection onto the wall; once small enough (spatial) dimensions are probed, 
the gravitational interaction will again appear large.  Models of this sort 
can remove the hierarchy problem, by eliminating the large difference in 
scales between the electroweak scale and the Planck mass.  An application of 
Gauss' law yields the result~\cite{ADD1}
\begin{equation}
M_{Planck}^2 \sim r^n M_{eff}^{2+n}
\end{equation}
where $r$ is the spatial size of the extra dimensions in the bulk, and 
$M_{eff}$ is the effective Planck mass.

Explicit suggestions have been made~\cite{other} for how such a low mass 
effective Planck or string scale and large extra dimensions might arise in 
both Kaluza Klein models and string theory.  We will concentrate on one such 
scenario in which large extra-dimensional gravity is embedded into string 
models~\cite{string}, where the string scale, $M_S$, is identified with the 
effective Planck mass, $M_{eff}$.  One interesting consequence of this 
scenario is that a Kaluza Klein (KK) tower of massive gravitons can interact 
with the Standard Model (SM) fields on the wall.  This can lead to direct 
production of a graviton tower as well as virtual exchange of a graviton 
towers.  Direct production of a graviton tower produces a missing $p_{_T}$ 
type signal, while virtual exchange can lead to new, tree-level interactions 
and/or modifications to SM processes.  The Feynman rules for these new types 
of interactions have been developed, {\it e.g.}, in Ref.~\cite{feynman}, and 
many processes have been studied in 
$e^+ e^-$~\cite{mpp,ck,tgr,lee,AB-SS},
$e \gamma$~\cite{AB-SS,dav2,gps}, 
$\gamma \gamma$~\cite{tgr,AB-SS,ad,dav1,mps,gmps}, $e p$~\cite{tgr,mrs1} and 
hadron~\cite{mpp,AB-SS,jlh,gmr,mrs2,mrs3,bal,kingman,oscar} colliders.  
New contributions to 
standard model interactions can occur in almost any process involving photon 
production and/or exchange or other neutral current phenomena.  Additionally, 
Higgs production~\cite{tgr2,xhe}, precision electroweak observable 
analyses~\cite{ew} and astrophysical constraints~\cite{astro} have been 
considered.  Based on direct production analyses, the current limits on 
$M_S$ fall in the range $500 \; GeV$ to $1.2 \; TeV$, while virtual graviton 
tower effects can yield current $M_S$ estimates from $650 \; GeV$ to 
$1.2 \; TeV$.  Future colliders, like the NLC and LHC can push these limits 
into the multi-$TeV$ range.

In this note, we will focus on aspects of dijet production at $\gamma \gamma$ 
colliders.  Other two-photon processes are also valuable in probing low-scale 
gravity effects~\cite{tgr,ad,dav1,mps,AB-SS,gmps}, but dijet production 
will be one of the most experimentally accessible processes in $\gamma \gamma$ 
collisions with guaranteed large event rates. The authors of 
Ref.~\cite{AB-SS} have 
recently considered gauge boson-gauge boson scattering in general, 
incorporating  the effects of low-scale gravity models, and include useful 
results for $\gamma + \gamma \rightarrow g + g$ which is necessary for our 
calculation. We also require, however, cross-sections for the corresponding 
$\gamma + \gamma \rightarrow q + \overline{q}$ processes for the two-jet 
cross-section at leading order.  The authors of Ref.~\cite{AB-SS} fail, 
however,  to include the ``box'' diagram: $\gamma + \gamma \rightarrow g + g$ 
exists as a 1-loop diagram in the SM~\cite{box}.  Although the box diagram, in 
the SM, is not as important in $\gamma \gamma$ collisions as it is in hadron 
collisions, we include it here for completeness \cite{in_proof}.

The authors of Ref.~\cite{EHMM} consider the inverse process, di-photon 
production at hadron colliders.  These authors present the parton level 
processes for both $g + g \rightarrow \gamma + \gamma$ (including the box 
diagram) and $q + \bar{q} \rightarrow \gamma + \gamma$.  The subprocesses we 
consider here, $\gamma + \gamma \rightarrow g + g$ and 
$\gamma + \gamma \rightarrow q + \bar{q}$, are identical in form, and differ 
only by color factors, from those presented in Ref.~\cite{EHMM}.  We will not 
reproduce those expressions here, but focus instead on optimizing the 
sensitivity of the $\gamma + \gamma \rightarrow jet + jet$ process to new 
physics contributions. 

\begin{flushleft}
 {\large {\bf II.~Calculation and Results}}
\end{flushleft}
\vskip 0.5cm

To examine the  $\gamma \gamma \rightarrow j j$ process at a future collider, 
we assume a linear $e^+ e^-$ collider, with backscattered laser 
photons~\cite{laser} for the initial photon beams.  The physical process at 
leading order is a sum of two ``parton level''  subprocesses, 
$\gamma \gamma \rightarrow g g$ and $\gamma \gamma \rightarrow q \bar{q}$; 
furthermore, the subprocesses include SM contributions as well as 
extra-dimensional gravity (KK graviton tower exchange) contributions.  In the 
SM, the lowest-order Feynman diagram for $\gamma \gamma \rightarrow g g$ is 
the one-loop, box diagram.  Although nominally higher-order in the 
perturbative expansion, we include it, as well as its interference with the 
extra-dimensional gravity contribution as its contributions are known to be 
very important in the inverse process (two-photon production in hadron 
collisions.) 

The event rate at planned colliders, even considering the SM contribution 
alone, is significant.  With the addition of graviton tower exchange, the 
angular and energy distribution of events is altered.  The graviton tower 
exchange is essentially the s-channel exchange of a large number of gravitons, 
all with different masses.  This leads to an enhancement of the cross section 
at all invariant masses kinematically allowed; a consequence of this is that, 
for low enough $p_{_T}$, the SM contribution dominates while at higher 
$p_{_T}$ the contribution of graviton tower exchange dominates.  Furthermore, 
the exact value of $p_{_T}$ where graviton tower exchange becomes important 
depends strongly on the scale parameter, $M_S$.  These properties are 
illustrated in Figure~1.

In Figure~1, we show some typical results of our calculation.  First, we 
choose an $e^+ e^-$ collider with $\sqrt{s} = 500 \; GeV$ operating in 
$\gamma \gamma$ mode, where the $\gamma$ beams are generated by backscattering 
laser photons off the original lepton beams.  In order to simulate detector 
accpetances, we employ cuts on our simulated events: $p_{_T} > 10 \; GeV$ and 
$\theta_{lab} > 10^\circ$ from the beam pipe are required to observe a jet.  
Below, we refer to this choice of acceptance cuts as nominal.  In order to 
compare and contrast dijet production, we present the $p_{_T}$ distribution 
for purely SM production (dashed curve), as well as SM + KK graviton tower 
exchange for $n=4$, and $M_S = 1.0 \; TeV$ (solid curve) and 
$M_S = 2.0 \; TeV$ (dotdashed curve).  The deviation from SM occurs at larger 
$p_{_T}$ for larger $M_S$.  Any particular value of $M_S$ will have a value of 
the $p_{_T}$ cut which maximizes the deviation from SM in total cross section: 
\begin{equation}
\Delta = \frac{\sigma - \sigma_{SM}}{\delta \sigma}
\end{equation}
where $\delta \sigma$ is the statistical uncertainty in the actual cross 
section.  With the nominal acceptance cuts, though, we expect in excess of 
$10^6$ events per year (using typical planned luminosities), at each 
center-of-mass energy considered below.  Large event rates are thus possible 
even if rather severe cuts are applied.  Given the behavior of the 
extra-dimensional gravity contribution illustrated in Figure~1, sensitivity to 
deviations from the SM (especially at large $M_S$) can benefit from a large 
$p_{_T}$ cut, removing much of the cross section where the SM dominates.  

In order to find the optimal value of the $p_{_T}$ cut, we have used an 
iterative process.  We begin with the nominal acceptance cuts listed above, 
and searched for the highest value of $M_S$ which gave a significant deviation 
from the SM.  We defined ``significant deviation'' to be a $2 \sigma$ 
(statistical) deviation.  Then, we used that value of $M_S$, and varied the 
$p_{_T}$ cut in order to maximize the deviation from the SM; we replaced the 
original $p_{_T}$ cut with this new value.  This process is repeated until the 
values of the $p_{_T}$ cut and $M_S$ are stabilized.  This iterative process 
converges very rapidly and we have repeated this optimization process for each 
center-of-mass energy considered.

To obtain specific estimates of possible $M_S$ limits, we have considered a 
1~year run at center-of-mass energies given by  $500 \; GeV$, $1 \; TeV$, 
$1.5 \; TeV$ and $2 \; TeV$.  We take conservative values for the integrated 
luminosity: $50 \; fb^{-1}$ at the $500 \; GeV$ collider, and $200 \; fb^{-1}$ 
at the others.  Longer running times or more optimistic luminosity values will 
simply increase the search reach.

As seen in the expression for the ``parton level'' subprocesses in 
Ref.~\cite{EHMM}, the cross section depends on the number of dimensions in the 
bulk, $n$.  So, in addition to different values of the center-of-mass energy 
of the linear $e^+ e^-$ collider, we also consider 2 values of $n$: $n=4$ and 
$n=6$.  Our results are summarized in Table~I where achievable limits on $M_S$ 
are shown, as well as the optimum value of the $p_{_T}$ cut for each 
center-of-mass energy.  In addition, achievable limits on $M_S$ using a 
nominal $p_{_T}$ cut are shown for comparison.  The optimization of the 
$p_{_T}$ cut increases the $M_S$ limits by at least $700 \; GeV$; as expected, 
the optimization is more effective for larger center-of-mass energy.  

It is interesting to note that the value of the optimum $p_{_T}$ cut is, in 
all cases, approximately $46\%$ of the beam energy of the $e^+ e^-$ collider.  
In addition to maximizing the deviation from the SM, this large value for the 
$p_{_T}$ cut indicates a very nice signature for extra-dimensional gravity 
effects: an excess at extremely large $p_{_T}$.

\begin{flushleft}
 {\large {\bf III.~Conclusions}}
\end{flushleft}
\vskip 0.5cm

In conclusion, we have examined dijet production at $\gamma \gamma$ colliders, 
in order to study the effects of, and search potential for, large 
extra-dimensional gravity models.  We have included a full, tree-level 
calculation of $\gamma + \gamma \rightarrow q + \bar{q}$ (SM plus KK graviton 
tower exchange), and the 1-loop ``box'' diagram (SM) plus tree-level, KK 
graviton tower exchange for $\gamma + \gamma \rightarrow g + g$.  Furthermore, 
we maximized the string scale, $M_S$, reach by optimizing the $p_{_T}$ cut.

We found that a rather large $p_{_T}$ cut yielded the highest sensitivity to 
the string scale.  At a $500 \; GeV$ linear $e^+ e^-$ collider, operating in 
$\gamma \gamma$ mode, using a cut of $p_{_T} > 115 \; GeV$, dijet production 
will be sensitive to $M_S$ from $2.75 \; TeV$ ($n=6$) up to $3.24 \; TeV$ 
($n=4$).  These sensitivities are $600-700 \; GeV$ higher than they would be 
with a nominal $p_{_T}$ cut of $10 \; GeV$.  At a $2 \; TeV$ linear $e^+ e^-$ 
collider, operating in $\gamma \gamma$ mode, using a cut of 
$p_{_T} > 465 \; GeV$, dijet production will be sensitive to $M_S$ from 
$9.35 \; TeV$ ($n=6$) up to $11.10 \; TeV$ ($n=4$).  At this higher 
center-of-mass energy, the increase in sensitivity, compared to the nominal 
$10 \; GeV$ $p_{_T}$ cut, is even more significant: $2.1-2.6 \; TeV$.  These 
limits assume a 1 year run at conservative luminosity estimates.  Longer runs 
or more optimistic luminosity estimates will, of course, increase the 
sensitivity to $M_S$ further.

Dijet production at $\gamma \gamma$ colliders is a sensitive and important 
test of large extra-dimensional gravity.  Although many other processes are 
also very sensitive to deviations from the SM as produced by large 
extra-dimensional gravity, it is important to have as many independent tests 
as possible, in order to verify the source of the deviations and to study the 
models as completely as possible.

\begin{flushleft}
 {\large {\bf IV.~Acknowledgments}}
\end{flushleft}
\vskip 0.5cm
The work of MAD was supported, in part, by the Commonwealth College of Penn 
State University under a Research Development Grant (RDG); the work of one of 
RR was supported, in part, by NSF grant DUE-9950702.

\newpage

{\Large
{\bf Figure Captions}}
\begin{itemize}
\item[Fig.\thinspace 1.] $p_{_T}$ distribution for dijet production at a $500 \; GeV$ 
$e^+ e^-$ collider operating in $\gamma \gamma$ mode.  The dashed curve 
indicates the SM cross section while the solid (dotdashed) curve indicates the 
contribution with the addition of extra dimension gravity with parameters 
$M_S = 1\, (2) \; TeV$ and $n = 4$.
\end{itemize}

\newpage

{\Large
{\bf Tables}}

\begin{table}[h]
\begin{center}
\begin{tabular}{|c|c|c|c|}
\hline
$\sqrt{s} \; (GeV)$ & $p_{_T}$ cut $(GeV)$ &  $M_S \ (GeV)$ &
$M_S \, (GeV)$ \\
&  & $(n=4)$ & $(n=6)$ \\ \hline
500  & 10  & 2500  & 2150 \\
500  & 115 & 3240  & 2750 \\ \hline
1000 & 10  & 4900  & 4000 \\
1000 & 230 & 6560  & 5700 \\ \hline
1500 & 10  & 6700  & 5700 \\
1500 & 350 & 8950  & 7500 \\ \hline
2000 & 10  & 8500  & 7200 \\
2000 & 465 & 11100 & 9350 \\ \hline
\end{tabular}
\end{center}
Table~I. $M_S$ limits possible with nominal and optimal $p_{_T}$ cut for  
$n=4$  and $n=6$
\end{table}


\begin{thebibliography}{99}
\bibitem{in_proof} 
As we were completing this analysis, we found Ref.~\cite{gmps}, which also 
calculates dijet production in $\gamma \gamma$ colliders.  These authors do 
include the $\gamma + \gamma \rightarrow q + \bar{q}$ subprocess, and so 
correct one of the flaws in Ref.~\cite{AB-SS}, however they also fail to 
include the box diagram.  In addition, as discussed below, the sensitivity 
of any $\gamma \gamma$ collider to the string scale $M_S$ can be increased 
with the application of a very severe $p_{_T}$ cut.  The authors of 
Ref.~\cite{gmps} discuss the sensitivity for various values of the $p_{_T}$ 
cut, but they do not systematically study this aspect.  In our analysis, we 
find optimal values for the $p_{_T}$ cut at each center-of-mass energy, 
yielding sensitivity to significantly higher $M_S$.
\end{thebibliography}
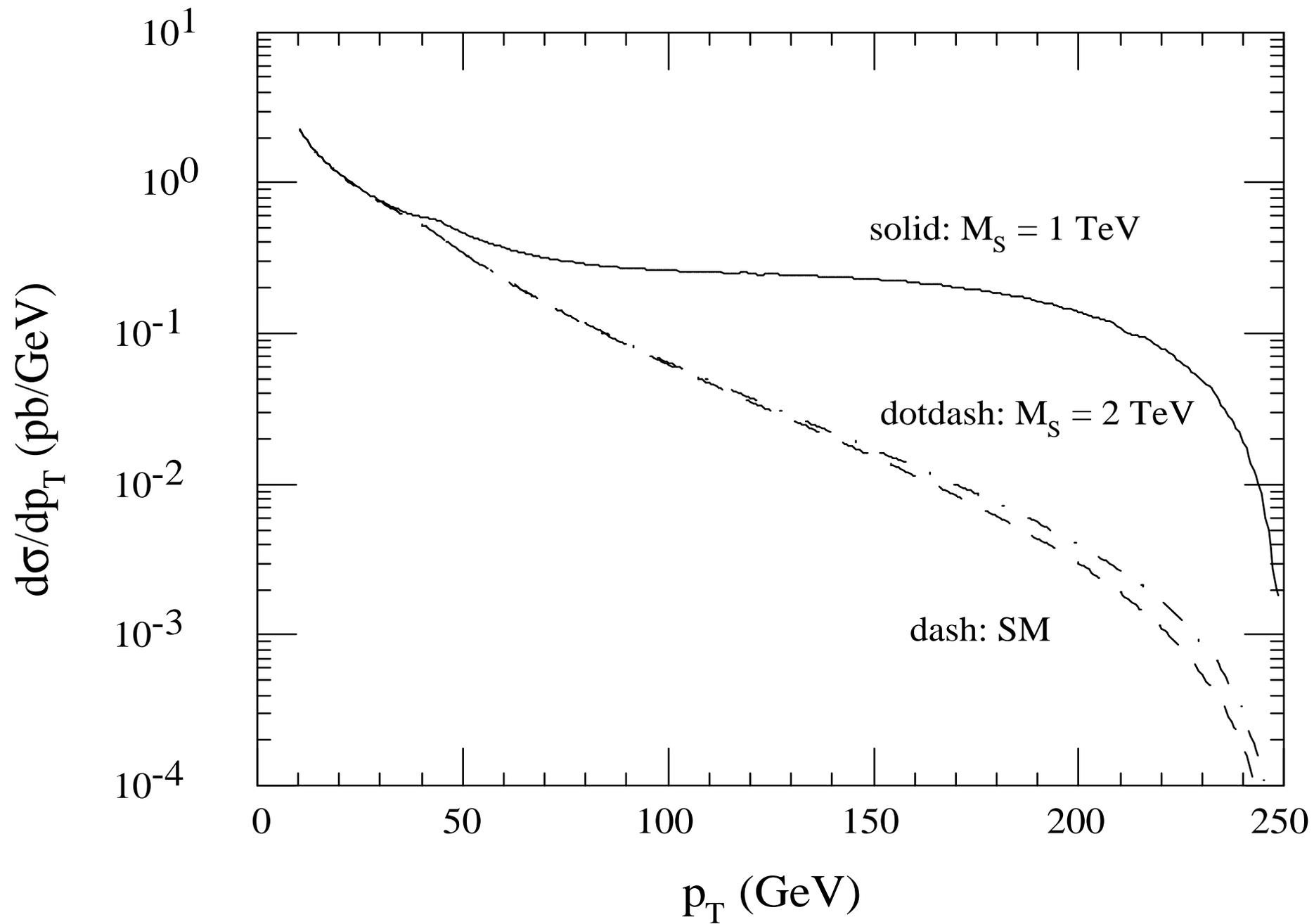
\end{document}